# Preparation, characterization and magnetic studies of $Bi_{0.5}X_{0.5}(X=Ca,Sr)MnO_3$ nanoparticles

S S Rao and S V Bhat*

*Department of Physics, Indian Institute of Science, Bangalore – 560012, India*

Nanoparticles (dia ~ 5 - 7 nm) of $Bi_{0.5}X_{0.5}(X=Ca,Sr)MnO_3$ are prepared by polymer assisted sol-gel method and characterized by various physico-chemical techniques. X-ray diffraction gives evidence for single phasic nature of the materials as well as their structures. Mono dispersed to a large extent, isolated nanoparticles are seen in the transmission electron micrographs. High resolution electron microscopy shows the crystalline nature of the nanoparticles. Superconducting quantum interferometer based magnetic measurements from 10K to 300K show that these nanomanganites retain the charge ordering nature unlike Pr and Nd based nanomanganites. The CO in Bi based manganites is thus found to be very robust consistent with the observation that magnetic field of the order of 130 T are necessary to melt the CO in these compounds. These results are supported by electron magnetic resonance measurements.

## Introduction:

The mixed valent perovskite manganites having the general formula $R_{1-x}A_xMnO_3$ (where R, rare earth ion like La, Pr, Nd and A is an alkaline earth ion like Ca, Ba, Pb, Sr) exhibit a rich variety of interesting properties like colossal magnetoresistance (CMR), charge ordering, orbital ordering and phase separation [1-5]. The properties of these manganites depend on the doping concentration (x), temperature, pressure, electric and magnetic fields [6-7]. Recently there have been reports on size dependent properties of these manganites [8-12]. But the Bi doped manganites show unusual properties in comparison with the rare earth based analogues though they (Bi and La) have similar ionic radii. The former exhibit high charge ordering up temperatures, ~ 525 K and are very stable even on the application of high magnetic fields. The charge ordering temperature of Bi based manganites increases with the increase of A site cationic radii unlike in the case of rare earth based manganites. The difference in the properties of these two types of manganites is attributed to the presence of highly polarizable $6s^2$ lone pair of electrons present on the Bi atom, and it is proposed that it strongly decreases the mobility of $e_g$ electrons and favors the charge ordering. Magnetic Fields of 130 T are needed to 'melt' the charge order and induce ferromagnetism and metallicity in these manganites. The Neel temperature $T_N$ and the charge ordering temperature $T_{CO}$ of $Bi_{0.5}Ca_{0.5}MnO_3$ are 120 K and 325 K and of $Bi_{0.5} Sr_{0.5}MnO_3$ are 110 K and 525 K[13-17] respectively. We have shown in our earlier work [8-9] that charge ordering is either fully or partly suppressed and there is a switch over from the anti-ferromagnetic phase to ferromagnetic phase, accompanied by a insulator to metal transition in Pr and Nd based nanomanganites. In this work, we show that even in ~ 5 nm sized nanoparticles of Bi based manganites the charge order is fully retained at least up to room temperature. In addition, the antiferromagnetic phase also seems to remain unaffected in these nanoparticles.

**Experimental:**

$Bi_{0.5}X_{0.5}(X=Ca,Sr)MnO_3$ nanoparticles are prepared by polymer precursor method[18]. As prepared samples are further heated (from 700°C to 900°C) to increase the particle size, but no increase in the particle size was observed. For convenience, these samples are labeled as BCMO7, BCMO9, BSMO7, BSMO9. X-ray diffraction (XRD), transmission electron microscopy (TEM), SQUID magnetometry and electron paramagnetic resonance (EPR) (Bruker X-band ESR spectrometer) measurements were used to characterize and to study the magnetic properties.

**Results and discussion:**

From the X-ray diffraction patterns as shown in the figure 1a and figure 1b, BCMO nanoparticles crystallize in monoclinic structure and BSMO nanoparticles crystallize in orthorhombic structure [16,19]. TEM measurements (figure 2a and 2b) on these samples have shown that the particles size is 5 nm and the size distribution is uniform. High-resolution electron microscopy (figure 2c) (HREM) shows that these particles are crystalline in nature. The interplanar spacing from the HREM picture is 3.84 $A^o$. These are the smallest sized nanomanganites ever reported. SQUID magnetometry measurements were done on all the four samples from 10 to 300 K at a magnetic field of 1 T.

Figure 3a shows the SQUID data of BCMO7 and BCMO9. It is seen that the qualitative behavior of these two samples is quite similar. They show anti ferromagnetic phase at around 120K and which is no different from the bulk value [14]. It is clear that the charge ordering nature of the compound doesn't change until room temperature except to say that there is a small rise in magnetization at around 270 K. It might be possible that at around room temperature, the nature of charge ordering is changing with the particle size. This has to be confirmed by other temperature variation measurements. The magnetization decreases with the particle size though the strength of the anti-ferromagnetic peak is seen to have decreased in BCMO7.

Figure 3b shows the SQUID data of BSMO7 and BSMO9 nanoparticles. These results show that the qualitative behavior of magnetization with temperature is similar. It is observed that the charge ordered phase is neither molten nor suppressed in the temperature range studied. There is no change in the anti-ferromagnetic transition the $T_N$ being 110 K which is the same as reported bulk value [14]. These observations are supported by EMR measurements and are shown in figure 4. These signals are very broad (line width is of the order of 2500 G) and weak in intensity, which are likely the signatures of charge ordering manganites [21]. If there is a size induced ferromagnetism, the line width of the EMR signals would be a few hundred gauss in the paramagnetic phase [20]. So it is most likely that the charge ordering phase of these nanoparticles is not changed until room temperature. Temperature variation (from 4K to 300K) EMR experiments are done on 5nm BCMO7 nanoparticles. DPPH (diphenyl picryl hydrazyl) is used as a standard field marker. As it is seen from the figure 4a, the EMR signals intensity decreases as one approaches towards anti ferromagnetic phase ($T_N$ = 120 K) from room temperature. Below 120 K, signal is disappeared which indicates the onset of ant-ferromagnetic phase. The sharp line centered around 3343 Oe is due to DPPH, used to find out resonance field accurately. The variation of EMR spectral parameters like line width ($8H_{PP}$) , resonance field (Ho) and the intensity (I) with the temperature are plotted in figure 4b. The line width increases with the decrease of temperature in the charge ordering phase. The resonance field and the intensity decrease as temperature approaches towards $T_N$. These properties give the finger print signatures of the charge ordering phase [21].

Temperature dependent neutron and electron diffraction experiments are underway to study the effect of particle size on the charge ordering phase. From these studies it is known that destabilizing the charge ordering state in Bi based manganites is difficult even by decreasing their size down to 5 nm, where quantum mechanical properties are expected to dominate. It seems that the hybridization between the 6s-Bi orbitals and 2p-O orbitals doesn't change with the particle size and hence the local distortion of $MnO_6$ octohedra and mobility of $e_g$ electron remains the same as it's bulk counterpart which favours charge ordering. From the above results it is known that the Bi based nanomanganites are quite different from rare earth based nanomanganites whose properties change dramatically with the particle size [8-10].

**Conclusions:**

Nanoparticles (of size 5 nm) of Bi based manganites were prepared using polymer precursor sol-gel method. These particles are characterized by XRD, TEM, SQUID and EMR spectroscopy. From the magnetization and EMR studies it is concluded that the reduction in the particle size has no effect on either the charge ordering phase or the anti ferromagnetic phase in the temperature range studied.

**Acknowledgements:** SSR would like to thank CSIR, Govt. of India for a senior research fellowship.

*For correspondence:
Email: svbhat@physics.iisc.ernet.in

**References:**

1. Colossal Magnetoresistance, Charge Ordering and Related Properties, eds. C.N.R. Rao, B. Raveau, World Scientific, Singapore, 1998.

2 . Colossal Magnetoresistive Oxides, ed., Y. Tokura G& B Science Publishers 2000

3. James C. Loudon, Neil D. Mathur & Paul A. Midgley, Nature **420**, 797, 2002.

4. H. Kawano-Furukawa and R. Kajimoto, H. Yoshizawa, Y. Tomioka, H. Kuwahara, Y. Tokura, Phys. Rev B, **67**, 174422 ,2003.

5. M. Fa¨th, S. Freisem, A. A. Menovsky, Y. Tomioka, J. Aarts, J. A. Mydosh, Science **285**, 1540 , 1999.

6. R. Senis, V. Laukhin, B. Martý´nez, J. Fontcuberta, X. Obradors, A. A. Arsenov and Y. M. Mukovskii, Phys.RevB, **57**, 14680, 1998.

7. Congwu Cui and Trevor A. Tyson, Phys.RevB, **70**, 094409, 2004.

8. S. S. Rao, K. N. Anuradha, S. Sarangi, and S. V. Bhat, Appl.Phys.Lett, **87**, 182503 ,2005

9. S. S. Rao and S. V. Bhat, Cond-mat 0512669.

10. M. A. Lopez-Quintela, L. E. Hueso, J. Rivas and F. Rivadulla. Nanotechnology **14**, 212, 2003.

11.A. Dutta, N. Gayathri and R. Ranganathan, Phys. Rev. **B 68**, 054432 ,2003.

12 . M. M. Savosta, V. N. Krivoruchko, I. A. Danilenko, V. Yu. Tarenkov, T. E. Konstantinova, A. V. Borodin, and V. N. Varyukhin, Phys. Rev B, **69**, 024413 , 2004.

13. Carlos Frontera , Jose´ Luis Garcý´a-Mun˜oz, Miguel ngel G. Aranda, Clemens Ritter, Anna Llobet, Marc Respaud and Johan Vanacken, Phys.Rev **B64**, 054401, 2001.

14. A. Kirste, M. Goiran, M. Respaud, J. Vanaken, J. M. Broto, H. Rakoto, M. von Ortenberg, C. Frontera,and J. L. Garcý´a-Mun˜oz, Phys.RevB, **67**, 134413 , 2003.

15. C Frontera, J L Garc´ýa-Mu˜noz, A Llobet, M A G Aranda, C RitterM Respaud and J Vanacken, J. Phys.: Condens. Matter **13**, 1071, 2001.

16. M. Hervleu, A. Maignan, C. Martin, N. Nguyen, and B. Raveau, Chem. Mater. **13**, 1356, 2001.

17. A. Trokiner, S. Verkhovskii, A. Yakubovskii, K. Kumagai, S-W. Cheong, D. Khomskii, Y. Furukawa, J. S. Ahn, A. Pogudin, V. Ogloblichev, A. Gerashenko, K. Mikhalev and Yu. Piskunov, Phys.RevB, **72**, 054442, 2005.

18. K.S. Shankar, S. Kar, G. N. Subbanna and A .K .Raychaudhuri, Sol. State. Comm. **129**, 479, 2004.

19. H. Woo, T. A. Tyson, M. Croft, S-W. Cheong and J. C. Woicik, Phys. Rev B, **63**, 134412, 2001.

20. F. Rivadulla , M. A. Lo´pez-Quintela, L. E. Hueso , J. Rivas, M. T. Causa, C. Ramos, R. D. Sa´nchez, and M. Tovar, Phys. Rev B, **60**, 11922, 1999.

21. Janhavi P. Joshi, Rajeev Gupta, A. K. Sood, S. V. Bhat, A. R. Raju and C. N. R. Rao, Phys. Rev B, **65**, 024410, 2001.

**Figure captions:**

1. Figures 1a and 1b show the XRD patterns of BCMO7 and BSMO7 nanoparticles.
2. TEM micrographs of (a) BCMO7, (b) BCMO9 nanoparticles. The scale bar in fig. 2a is 20 nm and in fig. 2b it is 5 nm. Figure 2c shows the HREM picture of BCMO7 single nanoparticle and the scale bar is 5 nm.
3. Magnetization behavior of (a) BCMO7 (squares) and BCMO9 (circles) and (b) BSMO7 (squares) and BSMO9 (circles) nanoparticles with temperature at a magnetic field of 1 T.
4. (a) EMR spectra at different temperatures. (b) Variation of EMR spectral parameters with temperature.

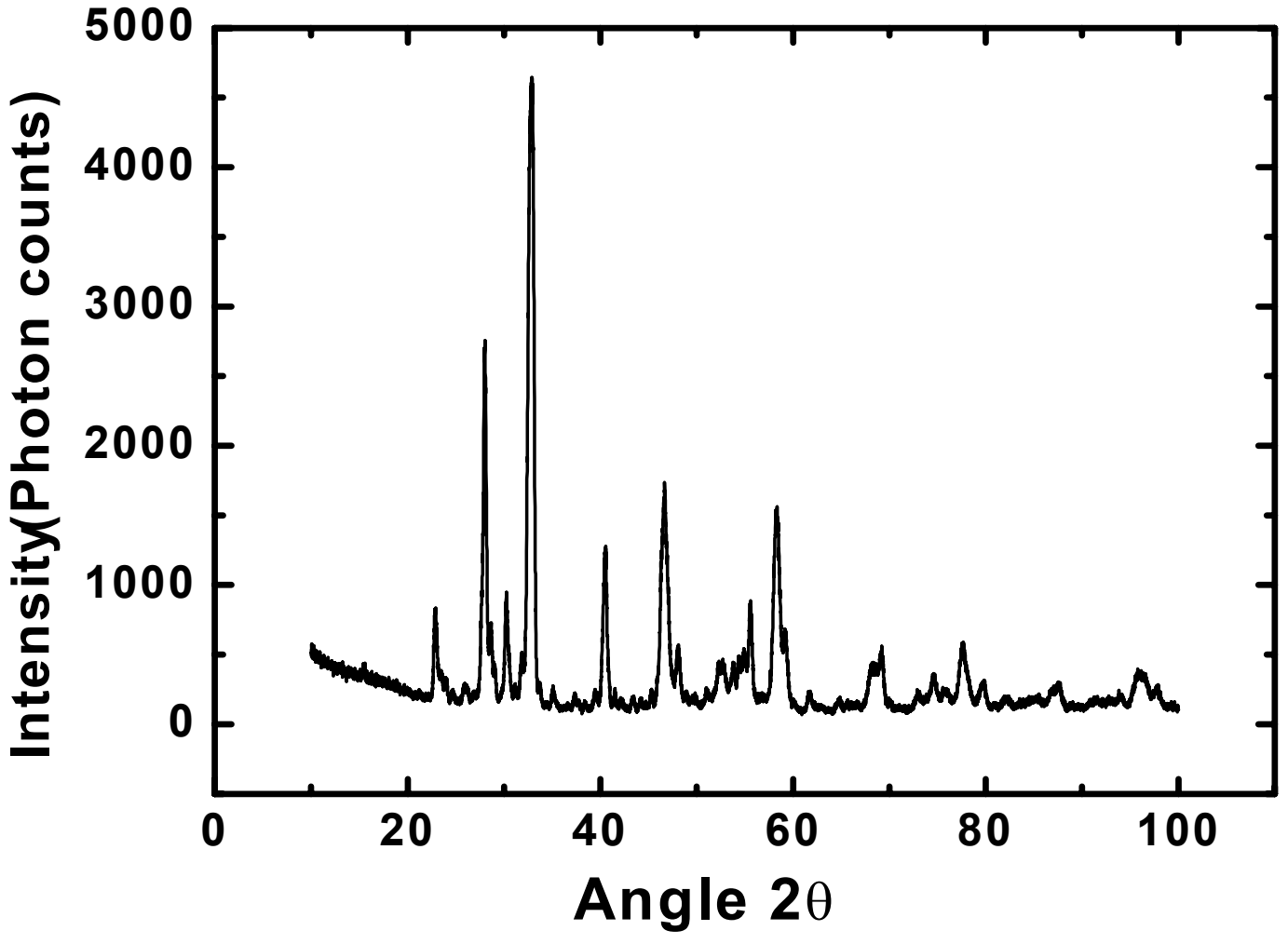

Figure 1a: Rao and Bhat

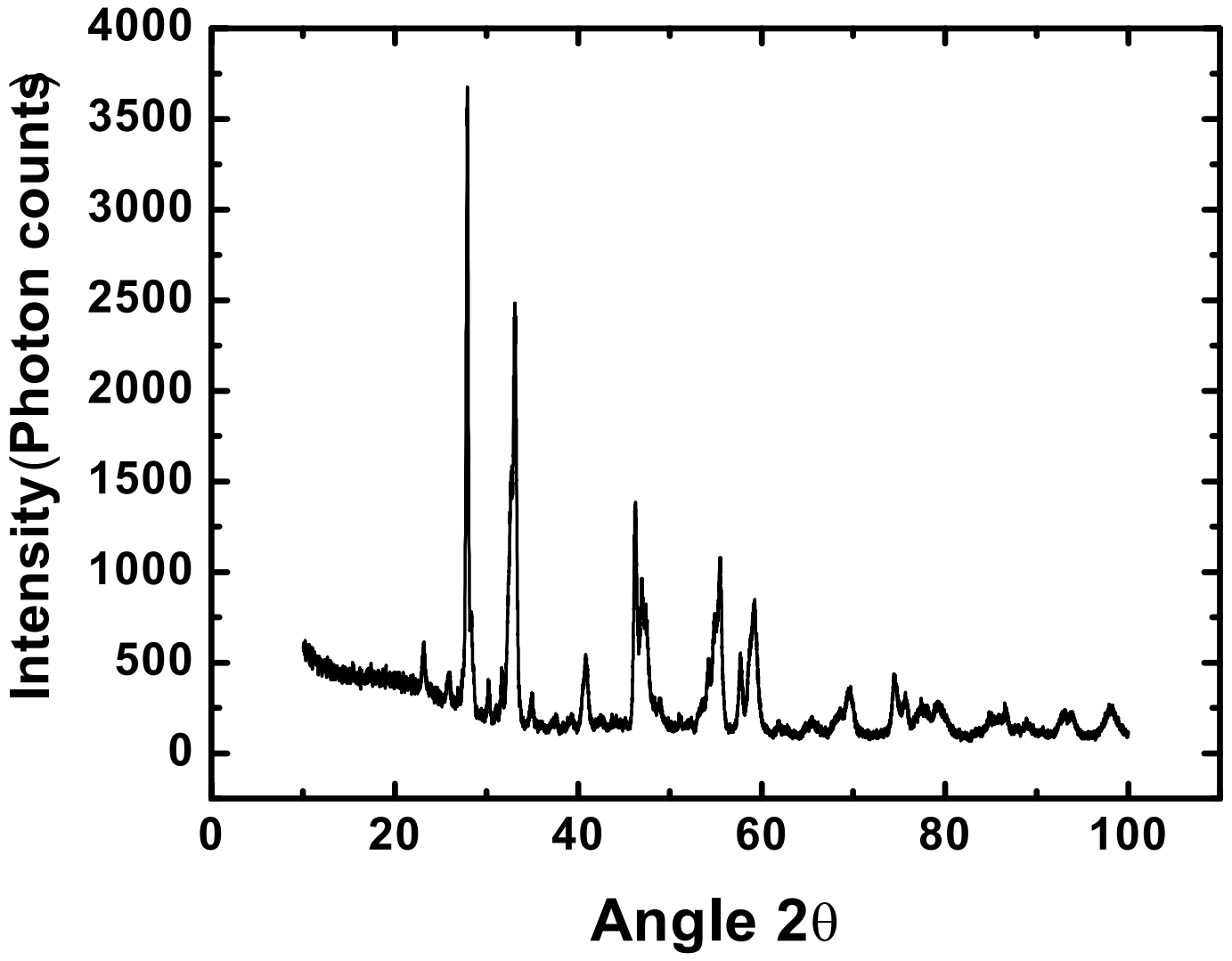

Figure 1b: Rao and Bhat

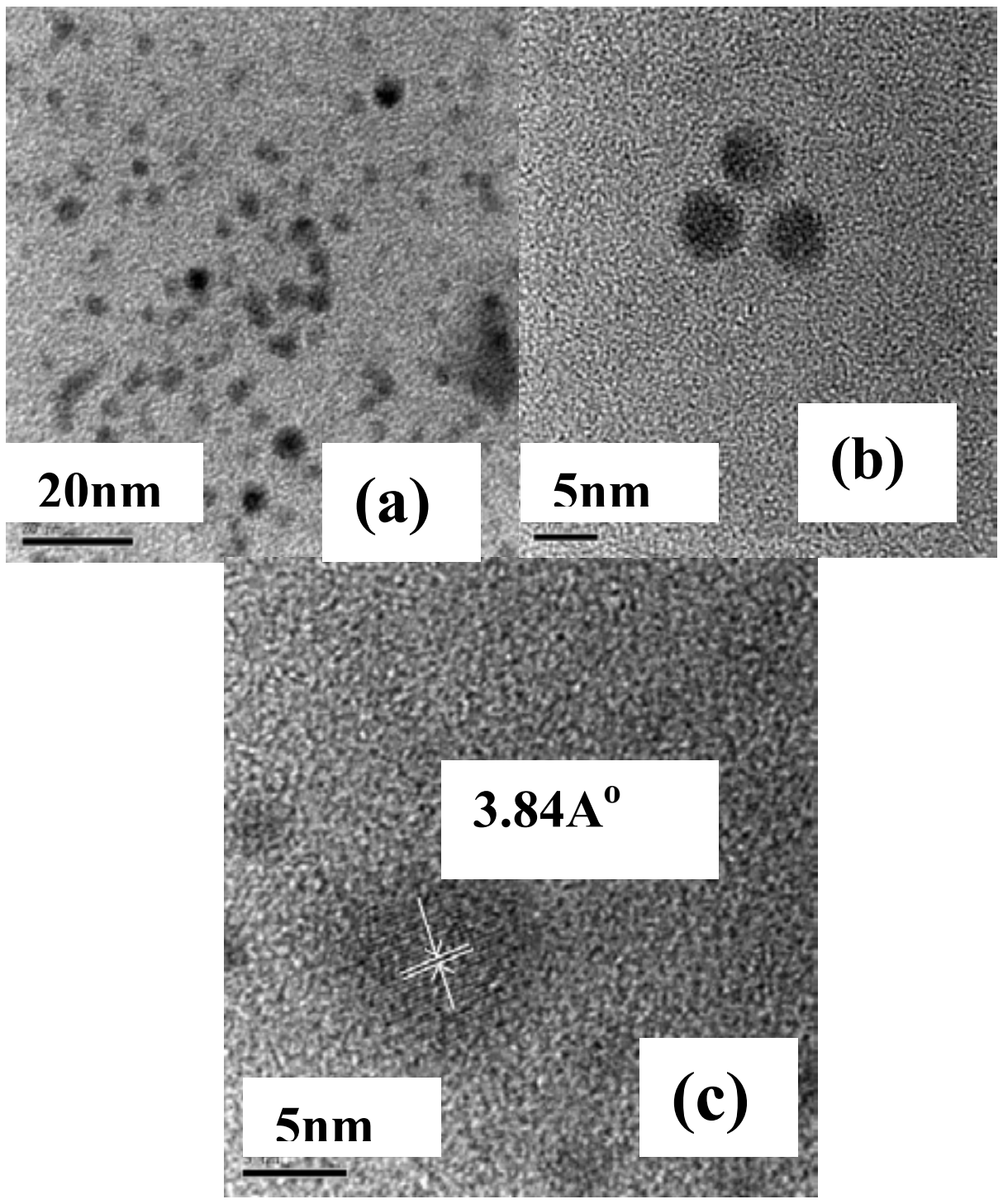

Figure 2: Rao and Bhat

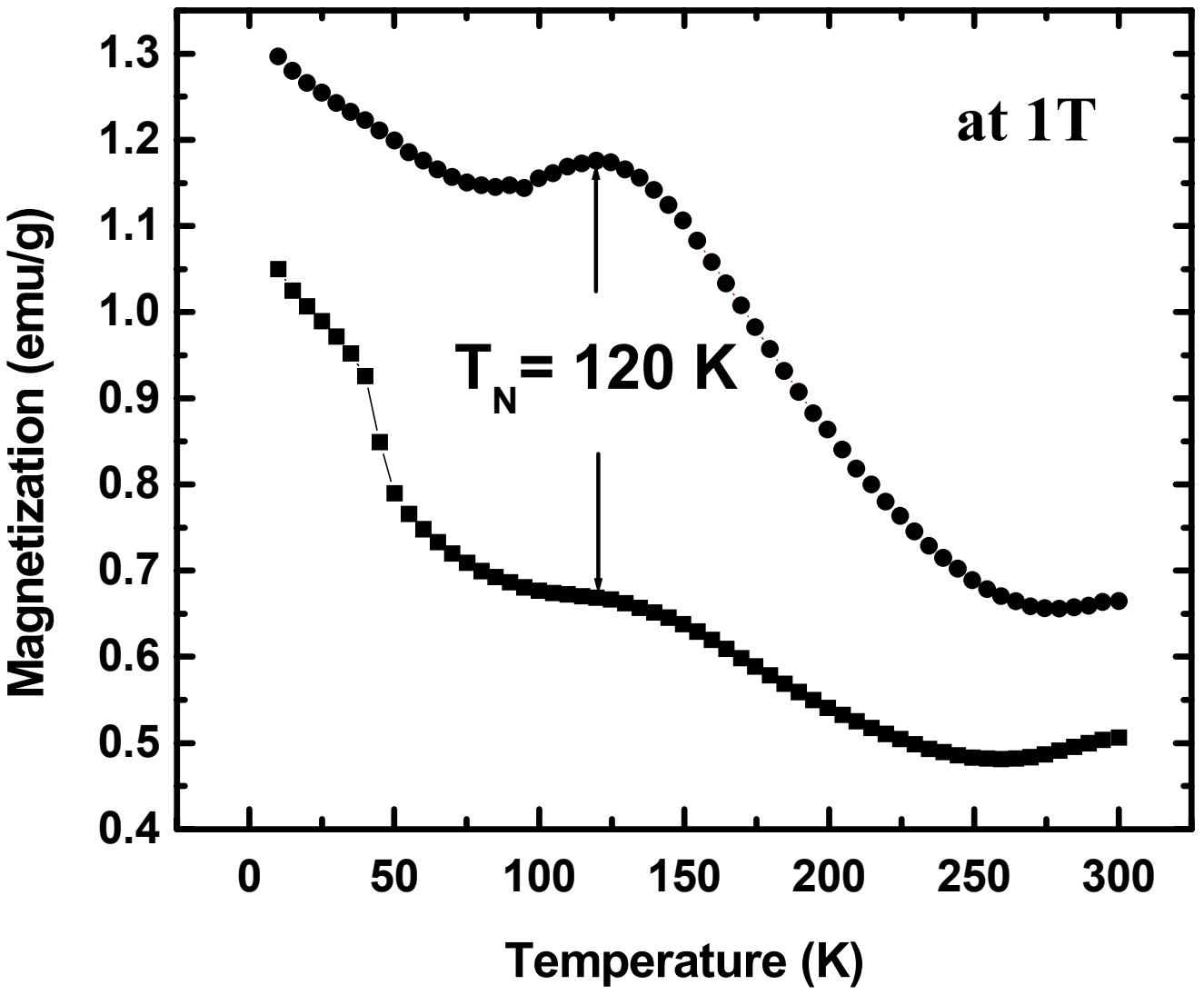

Figure 3a: Rao and Bhat

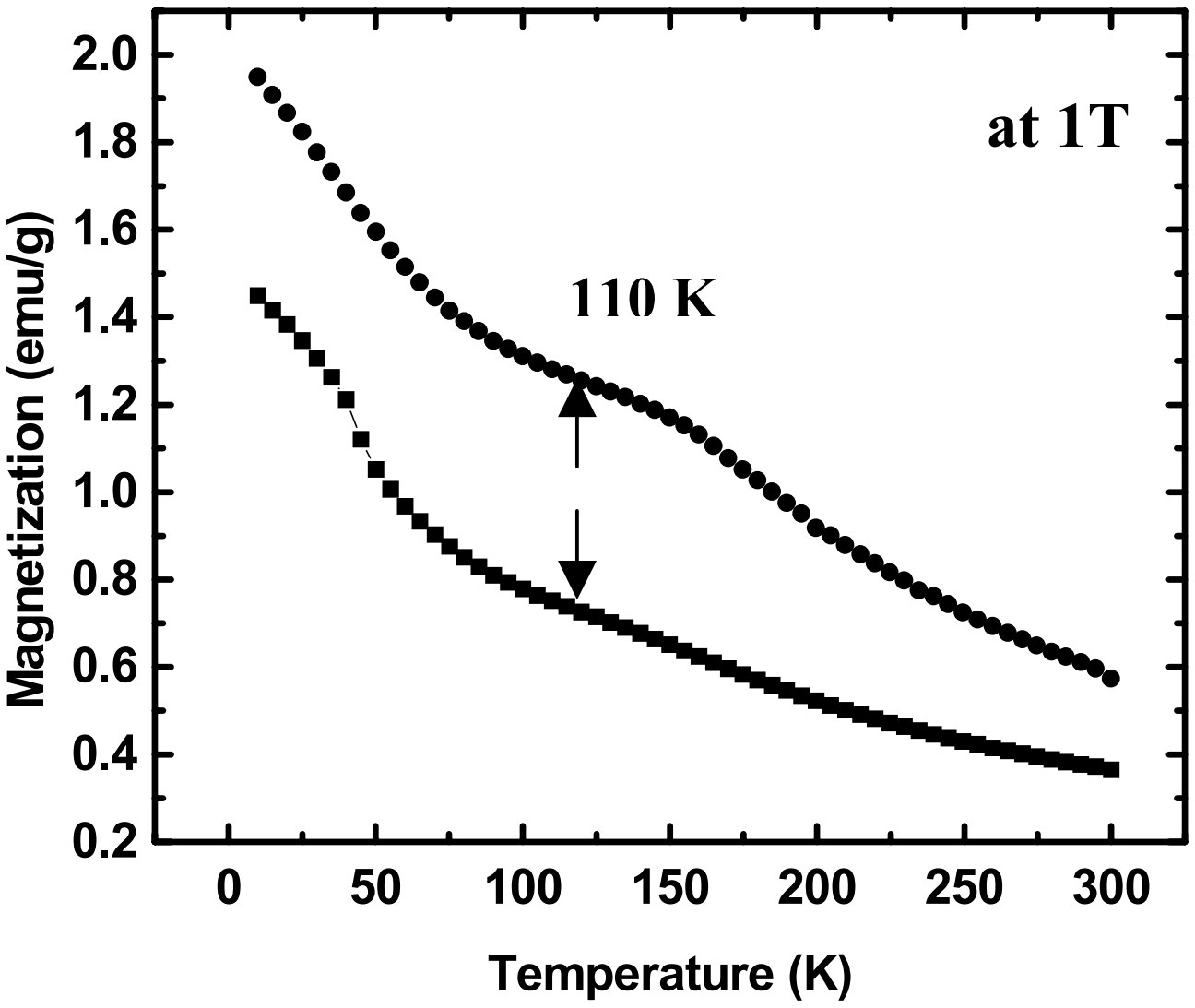

Figure 3b: Rao and Bhat

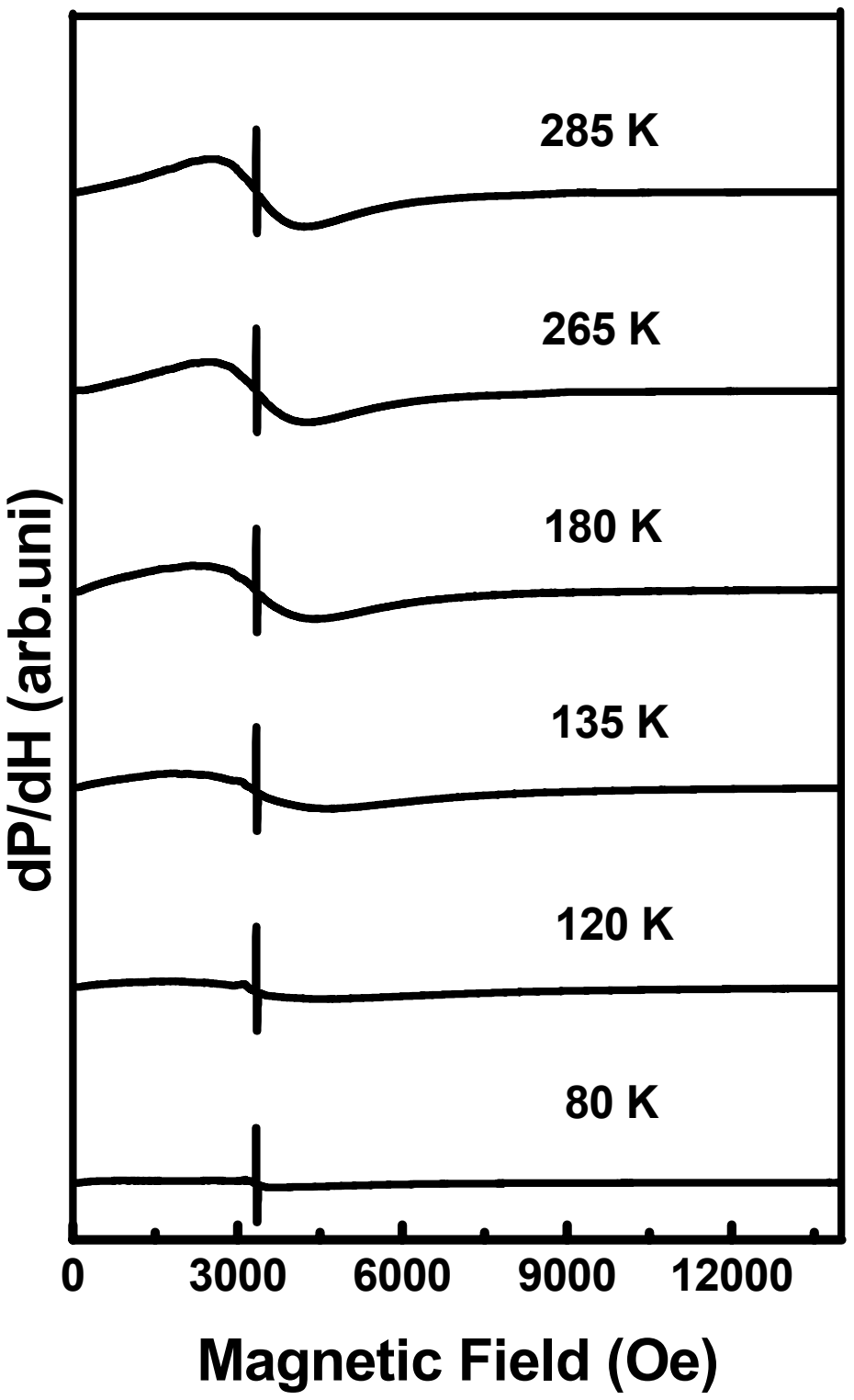

Figure 4a: Rao and Bhat

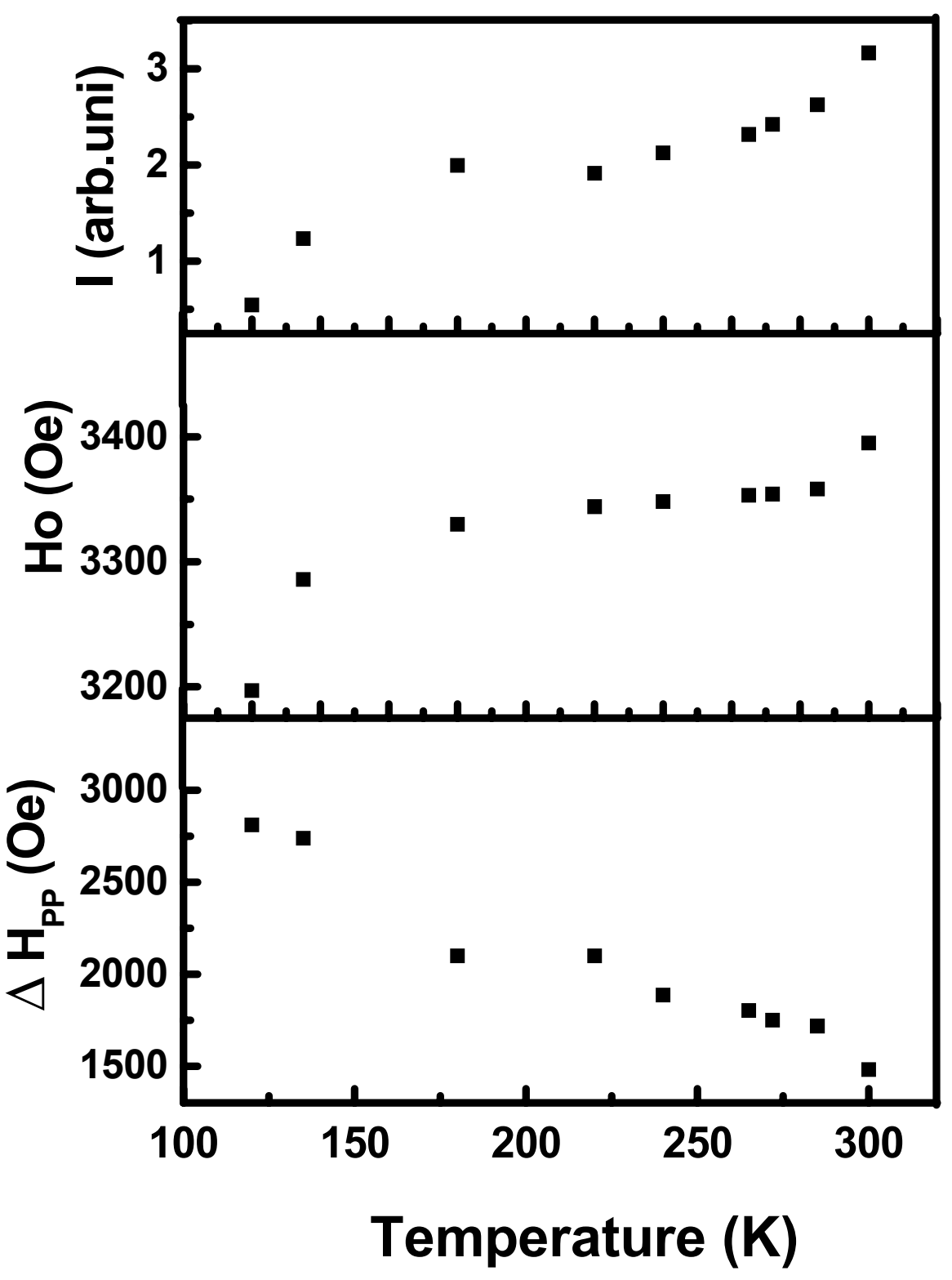

Figure 4b: Rao and Bhat